          [2004/06/22 1.2 (KOF); 2001/04/25 1.1 (PWD)]

\documentclass[a4paper,twocolumn]{Gaia2004} 
\usepackage{times}      
\usepackage{epsfig}     
\usepackage{natbib}     
\newcommand{\oversim}[2]{\protect{\mbox{\lower0.5ex\vbox{%
   \baselineskip=0pt\lineskip=0.2ex
   \ialign{$\mathsurround=0pt #1\hfil##\hfil$\crcr#2\crcr\sim\crcr}}}}}
\newcommand{\simgreat}{\mbox{$\,\mathrel{\mathpalette\oversim>}\,$}} 
\newcommand{\simless} {\mbox{$\,\mathrel{\mathpalette\oversim<}\,$}} 
\title{The Fundamental Building Blocks of Galaxies}
\author{Pavel Kroupa}
\affil{Sternwarte, University of Bonn, Auf dem H\"ugel 71, D-53121
  Bonn, Germany\\pavel@astro.uni-bonn.de}

\bibpunct{(}{)}{;}{a}{}{,}  

\begin{document}

\keywords{Methods: N-body simulations, Astrometry, Stars: formation,
  Galaxy: kinematics and dynamics, Galaxy: evolution, open clusters
  and associations: general}

\maketitle

\begin{abstract}
  Most stars form in compact, dense embedded clusters with memberships
  ranging from a dozen stars to many millions of stars.  Embedded
  clusters containing more than a few hundred stars also contain
  O~stars that disrupt the nebula abruptly. There are observational
  indications that the expulsion of gas from such clusters may occur
  within, or even quicker than, a stellar crossing-time scale.
  Because the star-formation efficiency is empirically deduced to be
  typically lower than 40~per cent, such clusters are left severely
  out of dynamical equilibrium with a super-virial velocity dispersion
  when they are exposed. As a consequence, they expand drastically.
  The velocity dispersion of expanding star-burst clusters may achieve
  values of a few tens~km/s, which would have a significant impact on
  the stellar velocity distribution function and therefore the
  morphology of a galaxy. An application of these ideas to the
  vertical structure of the Milky Way (MW) disk suggests that such
  popping star clusters may constitute the missing heat source for
  understanding the age--velocity-dispersion relation. If the early MW
  disk experienced a vigorous epoch of star formation then perhaps
  even the thick disk may have formed from popping clusters. Thus we
  have an alternative hypothesis to understanding thick disks which
  does not rely on, or may act in addition to, sinking galaxy
  satellites or sinking cosmological sub-structures.  The GAIA mission
  will test this hypothesis by allowing the kinematical fields around
  young clusters to be mapped to exquisite precision.
\footnote{To appear
  in the Proceedings of the Symposium "The Three-Dimensional Universe
  with Gaia", 4-7 October 2004, Observatoire de Paris-Meudon, France,
  eds: C. Turon, K.S.~O'Flaherty, M.A.C.~Perryman (ESA SP-576).}
\end{abstract}

\section{Introduction}
\label{sec:kr_intro}
Quoting from the review by Lada \& Lada (2003), ``observations of
nearby cloud complexes indicate that embedded clusters account for a
significant (70--90~\%) fraction of all stars formed in GMCs'' (giant
molecular clouds). This exciting discovery has deep implications for
the stellar distribution function in galaxies, and it is verified at
the other extreme of the star-formation scale, namely in violently
interacting gas-rich galaxies. de Grijs et al. (2003) find that star
cluster formation is a major mode of star generation in galaxy
interactions, ``with $>35~\%$ of the active star formation in
encounters occurring in star clusters''. Indeed, the entire Milky Way
(MW) population~II stellar spheroid could have been build-up from a
population of star clusters formed during the chaotic early merging
epoch of the Milky Way (MW), with the present-day globular clusters
forming the remaining high-mass end of the whole distribution (Kroupa
\& Boily 2002).

In order to assess the implications that the birth of stars in
clusters may have on galactic structure and kinematics, we need to get
a feeling for the physics of star cluster formation. This, however,
leads us rather rapidly to the limits of our present means, because
the formation of star clusters is intimately linked to stellar
feedback during the first few~$10^5$~yr. This is evident empirically
through the low star-formation efficiency in cluster-forming cloud
clumps of less than 40~per cent (Lada \& Lada 2003).  The theoretical
problem cannot be solved today because radiation transfer in a highly
dynamic, turbulent hydro-dynamical medium that contains complex
chemical networks and magnetic fields is untractable. Todays theory of
star-cluster formation is limited to purely hydrodynamic computations
(eg. Mac Low \& Klessen 2004; Bonnell, Bate \& Vine 2003) that give an
important insight into early fragmentation processes but cannot
capture the essential physics that act to expose the cluster. We
therefore need to resort to observations to constrain the important
physical processes responsible for exposing an embedded cluster.

Table~\ref{tab:kr_eg} provides a few well studied examples of very
young stellar systems, some of which have not yet evacuated their
remaining gas. Estimates of the crossing and relaxation times are also
indicated\footnote{The nominal crossing time is $t_{\rm cr} =
  2\,R_{\rm ecl} / \sigma_{\rm ecl} \approx 4\,\left(100\,M_\odot / M_{\rm
      ecl}\right)^{1/2} \; \left( R_{\rm ecl} / {\rm
      pc}\right)^{3/2}$~Myr; the relaxation time is $t_{\rm rel} =
  0.1\; N / {\rm ln}N\;t_{\rm cr}$. Here $R_{\rm ecl}$ is the
  characteristic radius and $\sigma_{\rm ecl}$ is the stellar 3D
  velocity dispersion; $N, M_{\rm ecl}$ are the number of stars and
  the stellar mass in the cluster.}, and it can be seen that $t_{\rm
  cr}>\;$age for the embedded low-mass objects that do not contain
massive stars, while the opposite is true for the rich clusters. The
massive clusters are well mixed ($t_{\rm cr}<\,$age) but not yet
relaxed, so some information from their formation is still evident.
\begin{small}
\begin{table}
\begin{center}
\leavevmode
\begin{tabular}{cccccc}
\hline\\[-2mm]
cluster   &age &$N$     &$R$
&$t_{\rm rel}$  &$t_{\rm cr}$\\
          &[Myr]  &        &[pc]  &[Myr]    &[Myr]\\[2mm]


\hline\\[1mm]

T--A (emb)  &1 &30 &0.5 &3.6  &4.1\\

$\rho$~Oph (emb)
&1          &100   &0.4 &3.5  &1.6\\

ONC (evac)
&1          &5000  &0.5 &18 &0.3\\

R136 (evac)
&2          &$10^5$ &2 &520  &0.6\\

Ant. (evac)
&$<10$      &$10^6$ &4  &3600 &0.5\\[1mm]

\hline
\end{tabular}
\end{center}
\vspace{-5mm}
\caption{\small
A few well-studied examples of very young clusters:
Taurus--Auriga (T--A) constitutes perhaps the smallest unit of
clustered star-formation and is heavily sub-clustered
(Brice{\~ n}o et al. 1998). 
$\rho$~Oph again is a low-mass cluster that contains no massive stars
and which is also sub-clustered (Bontemps et al. 2001). 
The Orion Nebula Cluster (ONC) has already evacuated its gas 
and contains no sub-structure 
(Wilson et al. 1997; 
Hillenbrand \& Hartmann 1998; Scally \& Clarke 2002). R136 is
the central cluster in the 30~Dor star-forming region in the Large
Magellanic Cloud (Brandl et al. 1996; Bosch et al. 2001). It already 
appears to have blown out most of
its gas, a large part of the cluster having been exposed. 
It contains no significant sub-clustering.
The star-burst clusters in the interacting
Antennae (Ant.) galaxies have typical radii $\approx4$~pc (Whitmore et
al. 1999). $t_{\rm cr}$ and $t_{\rm rel}$ are estimated using $R_{\rm
  ecl}=R$
and an average stellar mass of $0.4\,M_\odot$.
}
\label{tab:kr_eg}
\end{table}
\end{small}
Note that the evacuated clusters are likely expanding due to recent
gas blow-out (Section~\ref{sec:kr_gasexp}) and so $R>R_{\rm ecl}$
probably.

The important lesson from this table is that, while the radii of the
objects are always similar, their masses vary by about 5 orders of
magnitude.  In order to make first steps towards understanding the
implications star-cluster birth may have on galactic scales we
therefore need to worry about the distribution of $M_{\rm ecl}$;
$R_{\rm ecl}$ can be assumed constant for now.

\section{The birth of a cluster and its implication}
\label{sec:kr_birth}

\subsection{The embedded cluster}
\label{sec:kr_ecl}
Assume a region of a molecular cloud becomes gravitationally unstable
and contracts.  Contraction appears to occur very rapidly, on a
time-scale of 1--3~Myr for a wide range of cloud masses and sizes,
(Hartmann, Ballesteros-Paredes \& Bergin 2001). The free-fall collapse
time, $t_{\rm ff} = t_{\rm cr}/2^{3/2}$, for a clump with a radius of
5~pc and a mass of $10^3 (10^5)\,M_\odot$ is about 6~(0.6)~Myr.  As
the cloud clump contracts it fragments because the increasing density
leads to smaller Jeans masses, $M_{\rm J}\propto 1/\sqrt{\rho}$.
Turbulent motions also lead to localised fragments (Mac Low \& Klessen
2004). Within the fragments proto-stars form on a time-scale of
$10^5$~yr (Wuchterl \& Tscharnuter 2003) and decouple from the
hydro-dynamical flow thereby becoming ballistic.  They consequently
orbit within the contracting overall cloud clump.  Age-dating of the
young stars in the ONC, for example, suggests that the stars formed
over a time-span of about 1~Myr (Hillenbrand 1997; Palla \& Stahler
1999; O'dell 2001; Hartmann 2003), and so most of the stars have
orbited the forming embedded cluster a few times before star formation
was halted, probably due to the emergence of the O~stars. Indeed, the
short photo-evaporation time-scale of proplyds suggests that the
central star $\theta$1~C Ori with a mass of about $50\,M_\odot$ may
only be a few $10^4$~yr old (O'dell 2001).  A similar picture emerges
for the R136 cluster (Massey \& Hunter 1998; Bosch et al. 2001).

Because most of the stars have orbited a few times until that critical
time when the feedback energy from the massive stars ionises the gas
throughout the cluster region (with a typical radius of a few~pc)
thereby mostly halting further star formation, the embedded cluster
must be, at that instant, close to dynamical equilibrium. The stellar
velocity dispersion is
\begin{eqnarray}
\sigma_{\rm ecl} &=& \left({G\,\left(
        M_{\rm ecl} + M_{\rm g}\right)
          \over R_{\rm ecl}}\right)^{1\over2},\\
       &=& \left({G\,M_{\rm ecl}
          \over \epsilon\;R_{\rm ecl}}\right)^{1\over2},
\label{eq:sigma}
\end{eqnarray}
where $\epsilon\equiv M_{\rm ecl}/(M_{\rm ecl}+M_{\rm g})\approx
20-40$~per cent is the measured star-formation efficiency (Lada \&
Lada 2003).  Now note that for
\begin{equation}
\sigma_{\rm ecl} = 40\;{\rm km/s}, \quad\quad
{M_{\rm ecl} \over \epsilon\;R_{\rm ecl}} = 10^{5.5}\,M_\odot\;{\rm pc}^{-1}. 
\end{equation}
Since $R_{\rm ecl}\approx 3$~pc while $\epsilon \approx 1/3$ we find
that an embedded cluster with the mass of a typical globular cluster
has a velocity dispersion characteristic of the thick disk of the MW,
and indeed of dwarf galaxies. {\it The notion thus emerges that the
  birth of massive clusters may have an important impact on the
  structure and internal kinematics of galaxies} (Kroupa 2002).
But this can only be true (i) if such large velocity dispersions are
actually observed in embedded clusters, and (ii) if the embedded
velocity dispersion can be carried into the galactic field while the
cluster emerges from its cloud core.

\subsection{Dynamical state just before gas expulsion}
Observations indicate that condition~(i) may well be
fulfilled:\\[-7mm]
\begin{tabbing}
ONC: \= age $\approx$ 1~Myr; $\;$ \= $\sigma = 4.3$~km/s 
\quad \= {\small $\pm0.5$~km/s}\\
{\small (Jones \& Walker 1988}:
{\small proper-motion study, 900~stars)}\\[1mm]
R136: \> age $\approx$ 2~Myr; \> $\sigma = 55$~km/s
\>{\small $\pm19$~km/s}\\
{\small (Bosch et al. 2001}:
{\small spectroscopic l.o.s. velocities,}\\ 
{\small 48~O and B~stars)}\\[-7mm]
\end{tabbing}
The velocity dispersion for R136 is the three-dimensional dispersion,
based on the Bosch et al. (2001) line-of-sight value $\sigma_{\rm los}
= 32\pm11$~km/s. The ONC (Wilson et al. 1997) and probably also R136
(Brandl et al. 1996) contain little gas and thus the measured 3D
velocity dispersion is probably smaller than the pre-gas expulsion
stellar velocity dispersion within the embedded clusters,
$\sigma<\sigma_{\rm ecl}$, since the clusters are probably expanding
unless the star-formation efficiency was 100~per cent ($\epsilon=1$).
Furthermore, for the ONC the velocity dispersion is only a lower limit
because the proper motion analysis removes radial and tangential
motions.  In both clusters the observed stellar mass is too small to
account for the measured velocity dispersion.  The measured
super-virial nature of the ONC has been a puzzle for some time (Jones
\& Walker 1988; Hillenbrand \& Hartmann 1998).  The astonishingly high
velocity dispersion in R136 is attributed by Bosch et al. (2001) to be
the result of the orbital motions of massive companions.  The sample
can be split into two mass bins each containing 24~stars. Quoting the
line-of-sight dispersions and the average mass, \vspace{-4mm}
\begin{tabbing}
for $m>23.5\,M_\odot,\quad$ \= $\overline{m}=49.6\,M_\odot,\quad$
\= $\sigma_1=28$~km/s, \\
for $m<23.5\,M_\odot,\quad$ \= $\overline{m}=19.4\,M_\odot,\quad$
\= $\sigma_2=37$~km/s. 
\end{tabbing}
\vspace{-4mm} If the large $\sigma$ were due to orbital motions then
the more massive sample ought to have a larger velocity dispersion
since on average the binaries have more binding energy. Instead
$\sigma_1<\sigma_2$ which indicates that R136 is mass segregated with
a kinematically decoupled core of massive stars. Mass segregation is
indeed evident in star-counts of the cluster (Selman et al. 1999), and
in a mass-segregated cluster the massive stars are expected to have a
lower velocity dispersion.  Therefore probably a substantial part of
$\sigma$ may be due to the motions of stars through the cluster.

Both clusters thus appear to be well out of equilibrium and therefore
expanding.  Observations of the velocities of stars of later spectral
type in R136, and improved proper-motion studies of the ONC, are
needed to further test this notion. 

\subsection{Gas expulsion}
\label{sec:kr_gasexp}
Concerning condition (ii), $\sigma_{\rm ecl}$ will be conserved to
some degree {\it if} the star-formation efficiency is small ($\epsilon
\simless 40$~per cent) {\it and} gas expulsion occurs on a dynamical
time or shorter ($\tau _{\rm g} \simless t_{\rm cr}$). This is not
easy to verify empirically because we typically do not catch the
cluster just at the time when it is expelling its gas, although the
$\simless 0.2$~Myr old Treasure Chest cluster appears to be doing just
that.  Smith, Stassun \& Bally (2004) find the Treasure Chest cluster
to be very compact with a radius of less than 1~pc.  The HII region is
expanding with a velocity of approximately 12~km/s such that the
cluster volume will be excavated within a few~0.1~Myr.  Furthermore,
the star-burst clusters in the Antennae have $t_{\rm
  cr}\approx0.5$~Myr and Whitmore et al.  (1999) and Zhang, Fall \& Whitmore
(2001) find gas outflow velocities of 25--30~km/s. This corresponds to
a gas evacuation time-scale of 0.2~Myr which is comparable to $t_{\rm
  cr}$.  The star-burst clusters in the Antennae galaxies would thus
appear to support explosive gas expulsion.  Taking the evidence from
the ONC and R136 into account, it appears that {\it in the presence of
  O~stars explosive gas expulsion may drive early cluster evolution
  independently of cluster mass}.

Another handle on the evacuation time-scale can be obtained by
considering the energy input from massive stars and comparing this to
the binding energy, $E_{\rm bin}$, of the nebula. We have
\vspace{-3mm}
\begin{tabbing}
$|E_{\rm bin}|$ \= $=
 {G\,M_{\rm ecl+gas}^2 \over R_{\rm ecl}} = 8.6\times 10^{40} \;
\left({M_{\rm ecl+gas}\over M_\odot}\right)^2$~erg,\\
$t_{\rm cr}$ \> $= 4.8 \left({100\,M_\odot \over M_{\rm
ecl+gas}}\right)^{1\over2}\, \left({R_{\rm ecl} \over {\rm
pc}}\right)^{3\over2}$,
\end{tabbing}
\vspace{-3mm} and some characteristic values are listed in
Table~\ref{tab:kr_bind} for $R_{\rm ecl}=1$~pc.
\begin{table}
\begin{tabular}{ccc}
\hline\\[-2mm]
$M_{\rm ecl+gas}$ &$|E_{\rm bin}|$   &$t_{\rm cr}$\\[1mm]
$[M_\odot]$       &[erg]           &[Myr]\\[1mm]
\hline\\[-2mm]
$10^4$  &$8.6\times 10^{48}$  &0.48 \\
$10^5$  &$8.6\times 10^{50}$  &0.15 \\[0mm]
\hline
\end{tabular}
\caption{\small Binding energy and crossing time of two characteristic
  embedded clusters.}
\label{tab:kr_bind}
\end{table}

Resorting to Maeder's (1990) stellar-evolution tracks to evaluate the
mechanical (wind) and radiation energy output by massive stars we find
that a star with mass\\[-9mm]
\begin{tabbing}
$m = 15\,M_\odot \;$ \= injects $\;$ \= $3\times 10^{50}$~erg $\;$ \= 
per 0.1~Myr, and\\[0mm]
$m = 85\,M_\odot$ \> injects       \> $3\times 10^{51}$~erg  \> 
per 0.1~Myr.
\end{tabbing}
\vspace{-5mm}
The energy injected within a time shorter than the crossing time is
thus larger than the binding energy of the nebula, and {\it this supports
the conjecture that nebula disruption may typically occur within a
dynamical time-scale}, i.e very violently, explosively.

\subsection{Evolution of the exposed cluster}
\label{sec:kr_evol}

As a result of explosive expulsion of about 70~\% of a cluster's mass
the cluster stars are left orbiting with super-virial velocities and
expand nearly freely outwards. The classical energy argument suggests
that such clusters cannot survive (Hills 1980; Boily \& Kroupa 2003).
Basically, Pleiades- and Hyades-type open clusters ought not to exist,
unless the stellar IMF with which they were formed had no O~stars
(Elmegreen 1983; Kroupa, Aarseth \& Hurley 2001, hereinafter KAH).
This was a very interesting proposition, but subsequent observations
showed that clusters with $N\simgreat 500$ also contain massive stars;
evidence for a truncated IMF has never been found.

This problem of not being able to understand how open clusters form in
view of (1) explosive gas expulsion and (2) $0.2\simless
\epsilon\simless 0.4$, has recently been solved by the advent of
$N-$body work which treats the energy exchanges between stars due to
gravitational encounters precisely. Such work has been made possible
through theoretical research by Sverre Aarseth at Cambridge and Seppo
Mikkola at Turku on finding numerical and arithmetical solutions to
the non-linear gravitational dynamics problem (Aarseth
1999).\footnote{Recent work in this field has produced special-purpose
  gravity-pipe-line ({\sc Grape6}) supercomputers that can achieve a
  peak performance of 1~TFLOP per sec (Makino et al.  2003), while the
  next generation {\sc Grape8} machines are expected to reach a peak
  performance of the order of peta-FLOPS.}  Using an amended version
of the Aarseth integrator {\sc Nbody6} on a stand-alone PC, KAH
calculated the evolution of an initially dense embedded cluster
containing $10^4$ stars and brown dwarfs, with a 100~\% fraction of
binaries. The initial ONC model was taken to be the most likely
configuration from a parameter survey of all possible dynamical states
of the ONC (Kroupa, Petr \& McCaughrean 1999; Kroupa 2000).  Following
the seminal work by Lada, Margulis \& Dearborn (1984), KAH modelled
the gas component (initially 2/3rd of the total mass) as a
time-evolving background potential. This approximation makes the
calculations feasible on modern computers and is nicely consistent
with the much more CPU intensive SPH modelling that also needs to
treat radiative energy transfer in a highly simplified way (Geyer \&
Burkert 2001).  The KAH models include stellar-evolution-induced mass
loss and a solar-neighbourhood tidal field, and thus constitute the
most realistic existing computations of all relevant physical
processes acting during the emergence of a young cluster from its
natal cloud.

\begin{figure}[h]
\begin{center}
\leavevmode
\hspace{-20.01mm}\epsfig{file=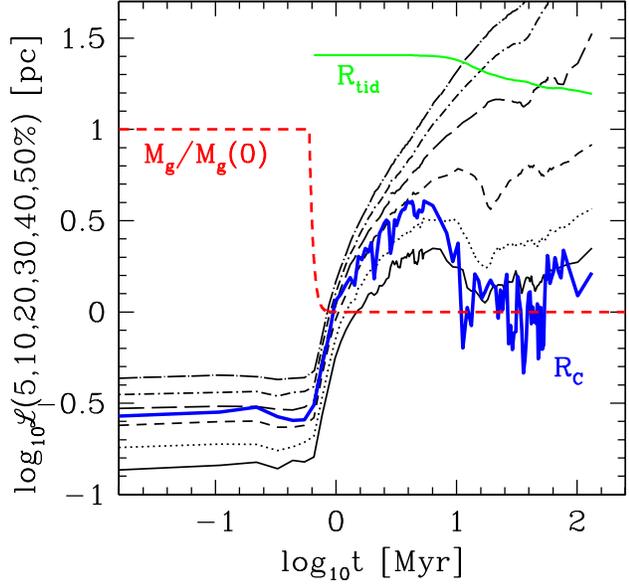,height=13cm,width=10cm,angle=0}
\end{center}
\vspace{-33mm}
\caption{\small The evolution of Lagrange radii (from bottom to top: radii
  containing 5, 10, 20...50 \% of the stellar mass), of the gas mass
  ($M_{\rm g}$), of the core radius ($R_{\rm C}$), and of the tidal
  radius ($R_{\rm tid}$), of an ONC-type cluster. The initial embedded
  model has a central density of $10^{4.8}$~stars/pc$^3$ with
  $\sigma_{\rm ecl}=6.8$~km/s, blows-out its gas after 0.6~Myr within
  a crossing time, and resembles the ONC by 1~Myr. After
  re-virialisation of about 1/3rd of the initial population the open
  cluster matches, by 100~Myr, the Pleiades (from KAH).}
  \label{fig:kr_lagr}
\end{figure}

The overall evolution of the embedded cluster with O~stars is
visualised in Fig.~\ref{fig:kr_lagr} which shows the reaction of the
mass shells to sudden gas removal. The Lagrange radii remain constant
during the embedded phase, but during the final stage just before gas
expulsion some mass segregation becomes apparent through a contraction
of the core radius.  Gas expulsion leads to expansion of the Lagrange
radii, and by 1~Myr the theoretical velocity dispersion has dropped to
the observed value in the ONC. The radial density profile also matches
the observed ONC by this time.  The 50~\% mass radius expands with a
velocity of about 1.2~km/s. It crosses the tidal radius by about 1~Myr
by which time about 50~\% of the embedded cluster mass becomes
unbound.  The inner $\approx 25$~per cent Lagrange radius as well as
the core radius contract after a few~Myr to reach a new minimum by
10~Myr.

A young open cluster has thus formed. It retains approximately $f_{\rm
  st}=1/3$rd of the population formed in the embedded cluster, and it
fills its Roche lobe entirely. By 100~Myr it evolves to an object that
resembles the Pleiades.
Theoretical cluster evolution models by Portegies Zwart et al. (2001)
furthermore show that the Pleiades will evolve to a Hyades-like
cluster, and so we are left to marvel at the amazing coincidence that
within less than one-third of a quadrant on the sky we can see the
following evolutionary sequence with the naked eye:
\vspace{-3mm}
\begin{tabbing}
\hspace{12mm} \= {\bf ONC} \= {$\mathbf\Longrightarrow$} \= {\bf Pleiades} 
\= {$\mathbf\Longrightarrow$} \= {\bf Hyades}\\

\>            \> \hspace{-4mm}  $\approx 99$~Myr      \>         
\> \hspace{-4mm} $\approx 500$~Myr       \>\\   
\end{tabbing}
\vspace{-8mm}
Much more work needs yet to be performed in order to better
understand the role of two-body relaxation and binary-star energy
exchanges during the critical gas-expulsion phase, and to quantify the
function 
\begin{equation}
f_{\rm st} = f_{\rm st}(M_{\rm ecl}, f_{\rm bin}, \epsilon, \tau_{\rm g}), 
\label{eq:kr_fst}
\end{equation}
where $f_{\rm bin}$ is the primordial binary proportion, and to
include the presence of the molecular cloud behind the ONC which is
likely to affect the kinematical field.  Nevertheless, these results
have essentially solved the open-cluster-formation problem.  {\it
  Bound clusters form readily despite a low star-formation efficiency
  and despite gas expulsion on a dynamical time-scale.}

\subsection{Predictions}
These results suggest that each cluster probably forms with about
three times as many stars than are contained within the later
re-virialised cluster.  There are two moving groups (MGs) associated
with each cluster, \vspace{-5mm}
\begin{tabbing}
MG I:$\;$ \=  the ``new'' MG stemming from explosive\\
         \> gas expulsion, and\\
MG II: $\;$ \> the classical MG stemming from cluster\\
         \> evaporation,
\end{tabbing}
\vspace{-5mm} such that the relative population of each is given
roughly, at some time $t$ after the formation time of the embedded
cluster, by\vspace{-5mm}
\begin{tabbing}
$\;N_{\rm MGII} \ll N_{\rm MGI}$, \hspace{5mm} \=
for \= $\;t     \ll \tau _{\rm ev}$,\\ 
$\;N_{\rm MGII}\approx (1/3) N_{\rm MGI}$, \>
for \> $\;t \simgreat \tau_{\rm ev}$,
\end{tabbing}
\vspace{-5mm} 
where 
\begin{equation}
\tau_{\rm ev}/[{\rm Myr}] \approx 10 (M/[M_\odot])^{0.75} 
\label{eq:kr_tev}
\end{equation}
(Baumgardt \& Makino 2003) is the evaporation or dissolution time of a
secularly evolving open cluster of mass $M$ on an approximately
circular Galactic orbit close to the solar radius.  The velocity
dispersion of MGII is $\sigma\approx 0$~km/s because the stars leak
out of the cluster with an energy close to zero (ignoring the rare
runaway stars that are shot out as a result of three- or four-body
encounters near the cluster core).  The velocity dispersion of MGI is
a function of the embedded configuration,
\begin{equation}
\sigma_{\rm MGI} = {\cal K}_{\rm th}(M_{\rm ecl}, \epsilon, \tau_{\rm
  g}) < \sigma_{\rm ecl},
\label{eq:kr_mgiiveldisp0}
\end{equation}
and this can be transformed to a correlation between the observed
velocity dispersion and the observed cluster mass, $M_{\rm cl}$, at
some time $t\ll\tau_{\rm ev}$,
\begin{equation}
\sigma_{\rm MGI} = {\cal K}_{\rm obs}(M_{\rm cl}),
\label{eq:kr_mgiiveldisp}
\end{equation}
which is expected to hold true for an ensemble of MGIs and for average
values of $\epsilon, \tau_{\rm g}$.  {\it We might thus be able to
constrain important star-formation parameters related to feedback
energy by studying eq.~\ref{eq:kr_mgiiveldisp0} theoretically and
eq.~\ref{eq:kr_mgiiveldisp} observationally}.

The theoretical research is time-consuming because the parameter space
spanned by $M_{\rm ecl}, R_{\rm ecl}, \epsilon, \tau_{\rm g}, f_{\rm
  bin}$ needs to be sampled using precise $N$-body calculations. For
$M_{\rm ecl}\simgreat 10^4\,M_\odot$ ($\simgreat 10^4$ stars) this
work will need the afore mentioned special-purpose {\sc Grape6}
supercomputers (Baumgardt \& Makino 2003).  The observational research
will need space-based all-sky astrometry missions that capture a
substantial fraction of local stars.

\subsection{The role of GAIA}

According to the above theory, each young cluster should be surrounded
by a radially expanding population of stars of similar age as the
cluster population. The expansion field will not be exactly
spherically symmetric because variations in the tidal field deflect
the stellar orbits. Nevertheless, a relation of the form of
eq.~\ref{eq:kr_mgiiveldisp} should emerge upon measurement of the
kinematical field surrounding young clusters.

The GAIA mission is ideally suited to uncover
relation~\ref{eq:kr_mgiiveldisp} locally in the MW and in the Large
and Small Magellanic Clouds (respectively LMC, SMC). Locally the high
precision of the proper motion measurements will allow highly detailed
mappings of all existing MGI and MGII. In the distant LMC only the
early-type stars can be measured around young clusters, but
nevertheless ${\cal K}_{\rm obs}$ ought to become evident.

The tangential velocity to the line-of-sight is
\begin{equation}
V_{\rm T}[{\rm km/s}] = 4.74 \; \times \; \mu[{\rm as/yr}] \; \times
\; d[{\rm pc}],
\label{eq:kr_vtan}
\end{equation}
where $\mu$ is the proper motion in arc sec per year, and $d$ is the
distance of the star. Neglecting pre-main sequence evolution for the
sake of illustration, an A~type star with $M_{\rm V}=1.1$ at the
distance of the LMC ($d=50$~kpc) has an apparent magnitude $m_{\rm
  V}=19.6$. From table~8 in Perryman et al. (2001) the measurement
accuracy for such a star corresponds to 28~km/s. This should uncover
radial flows away from massive young clusters if young stellar samples
can be combined in sub-regions around the clusters.  A G-type star
with $M_{\rm V}=5$ at a distance of the ONC (450~pc) has $m_{\rm
  V}=13.3$.  For such a star GAIA will have a precision of
0.011~km/s, which will suffice to define the kinematical state of
the entire ONC, although issues of crowding do arise.

\section{Creating a galaxy}

The ideas developed above can be explored further theoretically by
considering their implications on galaxy morphology.  In
Section~\ref{sec:kr_ecl} it was already noted that the birth of
massive clusters may impact the vertical structure of a disk galaxy
such as our own MW.  In order to study this in more detail it is
necessary to build a field population from an ensemble of popping
clusters.

\subsection{Basics}
The basic equation for the velocity distribution function of stars is
obtained by adding all expanding populations from all clusters formed
in one ``epoch''. Note that Weidner, Kroupa \& Larsen (2004) find that
the ECMF appears to be completely populated for star-forming epochs
lasting typically about 10~Myr, independently of the star-formation
rate (SFR).  For this purpose we need only to consider the
distribution of the clusters by mass formed in each epoch since the
radii of very young clusters do not differ significantly
(Section~\ref{sec:kr_intro}),
\begin{eqnarray}
&{\cal D}(v_z; M_{\rm ecl,max}, \beta) = \nonumber\\
&\hspace*{-4mm}\int\limits_{M_{\rm ecl,min}}^{M_{\rm icl,max}}
D(v_z; M_{\rm ecl})\, N(M_{\rm ecl}) \, \xi_{\rm
ecl}(M_{\rm ecl},\beta)\, dM_{\rm ecl},
\label{eq:kr_vdistr}
\end{eqnarray}
where ${\cal D}\,dv_z$ is the resulting number of stars with
$z$-components of their velocity vectors in the interval $v_z,
v_z+dv_z$, $M_{\rm ecl,min}\approx10\,M_\odot$, and $M_{\rm ecl, max}$
is the maximum embedded cluster mass in the cluster ensemble. The
product $N\,\xi_{\rm ecl}\,dM_{\rm ecl}$ is the number of stars formed
from clusters with masses in the range $M_{\rm ecl}, M_{\rm
  ecl}+dM_{\rm ecl}$. The expanding population of MGI stars emanating
from a cluster of mass $M_{\rm ecl}$ has a velocity distribution
function $D$, where $D(v_z; M_{\rm ecl})\,dv_z$ is the number fraction
of cluster stars with velocities in the range $v_z, v_z + dv_z$.  The
physics of star formation is contained in $D(v_z; M_{\rm ecl})$, but
for an assessment of the impact of clustered star formation on
vertical Galactic disk structure it suffices to consider a Gaussian
$v_z$ distribution, i.e.  a Maxwellian speed distribution in the
vertical direction, with a dispersion given by eq.~\ref{eq:sigma}.
The actual $D(v_z; M_{\rm ecl, max})$ would need to account for the
MGII, i.e. the fraction of stars that re-virialise in the bound
cluster, and which, upon secular evaporation from the cluster,
contribute a population of stars with a small velocity dispersion to
${\cal D}$.  Gravitational retardation would also need to be
incorporated; the stars that expand from a cluster are decelerated by
the mass within their distance from the cluster, such that
$\sigma_{\rm MGI}<\sigma_{\rm ecl}$.  These points are incorporated in
the exploratory investigation of Kroupa (2002). Before proceeding
further we need to specify the embedded cluster mass function.

\subsection{The mass function of embedded clusters}
As is apparent from eq.~\ref{eq:kr_vdistr} the fundamental
distribution function that governs ${\cal D}$ is the embedded cluster
mass function (ECMF).  Most empirical constraints suggest the ECMF to
be a power-law,
\begin{equation}
\xi_{\rm ecl}\propto M_{\rm ecl}^{-\beta}, 
\label{eq:kr_ecmf}
\end{equation}
$dN_{\rm ecl} = \xi_{\rm ecl}\,dM_{\rm ecl}$ being the number of
embedded clusters in the mass interval $M_{\rm ecl}, M_{\rm
  ecl}+dM_{\rm ecl}$.

Lada \& Lada (2003) find, for the local embedded cluster sample with
masses between about 20~and $10^3\,M_\odot$, that the ECMF can be
described as a power-law with $\beta\approx 2$, with a possible
flattening below roughly $50\,M_\odot$. The form of the ECMF below
about a few dozen~$M_\odot$ is very uncertain though because such
stellar groups are difficult to find. They disperse within a time
comparable to the gas removal time-scale (Adams \& Myers 2001; Kroupa
\& Bouvier 2003).

The maximum mass of a cluster, $M_{\rm ecl,max}$, in a freshly hatched
cluster ensemble correlates with the SFR of a galaxy (Weidner et al.
2004). In order to extend the ECMF to larger masses we therefore need
to study galaxies that have a sufficiently high SFR.  The Large and
Small Magellanic Clouds are such systems, and Hunter et al. (2003)
find $2\simless\beta\simless2.4$ for $10^{3} \simless M_{\rm
  ecl}/M_{\odot} \simless 10^{4}$.

The Antennae galaxies are composed of two currently merging major disk
galaxies, and the SFR in this system is sufficiently high to sample
the cluster MF to $10^{7-8}\,M_\odot$.  For a sample of $<10$~Myr old
clusters in the Antennae galaxies, Zhang \& Fall (1999) arrive at
$\beta = 1.95\pm0.03$ for clusters with masses in the interval $10^4
\simless M_{\rm ecl}/M_\odot \simless 10^6$ and ages $2.5<t/{\rm
  [Myr]}<6.3$.  ``Clusters'' more massive than about
$10^{6-7}\,M_\odot$ are observed to be composed of star-cluster
complexes (Kroupa 1998; Zhang et al. 2001), thus setting an empirical
maximum mass of about $10^{6-7}\,M_\odot$ for a true star cluster,
i.e. an equal-age, equal-metallicity population (Weidner et al. 2004).
Kroupa (1998) and Fellhauer \& Kroupa (2002a,b) propose such cluster
complexes to be the precursors to a number of recently found exotic
objects: ultra-compact dwarf galaxies and low-density clusters.

For a variety of environments the ECMF thus appears to have
\begin{equation}
\beta \approx 2.
\end{equation}
This is rather remarkable suggesting a possible universality of the
ECMF.

Note that, apart from the Lada\& Lada work, the above studies actually
report the shape of the mass function of young clusters rather than
embedded clusters. The studied cluster systems typically have ages
younger than 10~Myr. Most clusters younger than 10~Myr would still
contain the majority of their stars within a relatively compact region
even if they are expanding as a result of gas blow out
(Fig.~\ref{fig:kr_lagr}). Nevertheless, this is an issue worth keeping
in mind; the ECMF underlying the young-cluster systems of Hunter et
al. (2003) and Zhang \& Fall (1999) may well have a somewhat different
$\beta$. Kroupa \& Boily (2002) stress that as a result of the low
$\epsilon$ a distinction between the cluster mass function and the
ECMF is necessary, because clusters loose significant stellar mass as
a result of gas expulsion. Clearly, this is an important issue for
future $N$-body work.

\subsection{Thickening of galactic disks through clustered star formation}
\begin{figure}[h]
\begin{center}
\leavevmode
\centerline{\epsfig{file=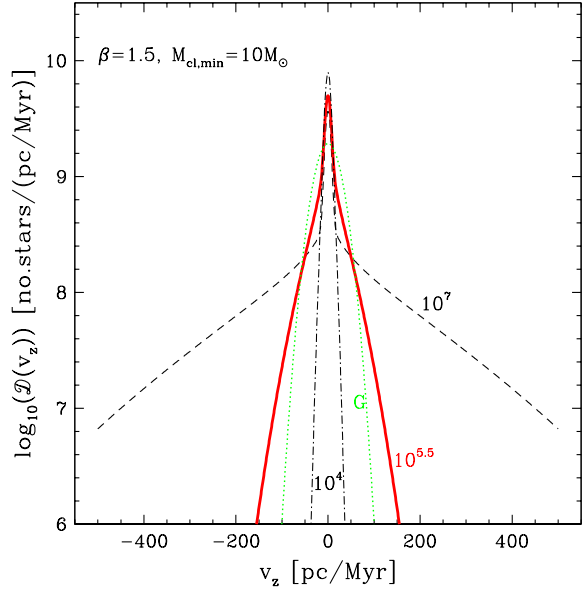,height=11cm,width=8cm,angle=0}}
\end{center}
\vspace{-25mm}
\caption{\small The resulting distribution function of vertical velocities 
  for a population of stars formed in one epoch of star formation
  which produced a power-law ECMF with $\beta=1.5$. The distribution
  function is shown for different maximum cluster masses, $M_{\rm
    ecl,max}=10^4, 10^{5.5}, 10^7\,M_\odot$. The curve labelled ``G''
  is a Gaussian distribution function with a velocity dispersion equal
  to the $M_{\rm ecl,max}=10^{5.5}\,M_\odot$ case (26~km/s).  The
  cluster--cluster velocity dispersion is 5~km/s.  Note that
  1~km/s~$\approx$~1~pc/Myr.  From Kroupa (2002).}
\label{fig:kr_vdistr}
\end{figure}

Given the ECMF and a pre-scribed $M_{\rm ecl,max}$,
eq.~\ref{eq:kr_vdistr} can be calculated. For $\beta=1.5$ and
different $M_{\rm ecl,max}$, the resulting ${\cal D}(v_z)$ is
non-Gaussian, having a ``cold'' peak with broad warm to hot wings
(Fig.~\ref{fig:kr_vdistr}). The width of the wings is determined by
$M_{\rm ecl,max}$, while the strength of the cold peak is given by
$M_{\rm ecl,min}$ and the cluster--cluster velocity dispersion. 
The distribution of the $z$-components of the velocity vectors of
solar-neighbourhood M~dwarfs indeed shows non-Gaussian features (Reid,
Hawley \& Gizis 1995) that are, at least qualitatively, consistent
with this scenario. Again, the GAIA mission will quantify the shape of
the velocity distribution of field stars allowing tests of the present
theory.

\begin{figure}[h]
\begin{center}
\leavevmode
\centerline{\epsfig{file=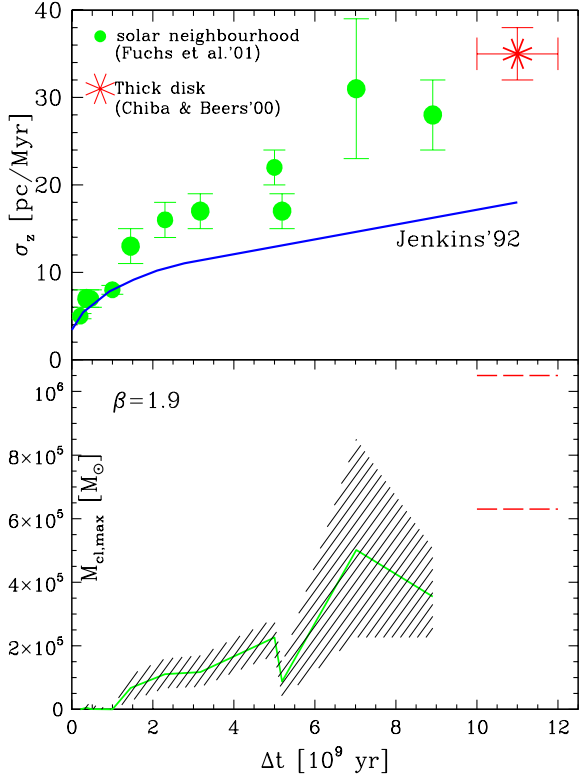,height=11cm,width=8cm,angle=0}}
\end{center}
\vspace{-8mm}
\caption{\small {\bf Upper panel:} The age--velocity-dispersion data for
  velocity components perpendicular to the MW disk for solar
  neighbourhood stars as compiled by Fuchs et al. (2001) are shown as
  solid dots. Jenkins' (1992) theoretical relation is plotted as the
  solid curve. The thick-disk datum is indicated by the old star.
  {\bf Lower panel:} The maximum cluster mass, $M_{\rm ecl,max}$, in
  the population of star clusters needed such that the difference in
  the top panel between observed data and theory vanishes is
  calculated and plotted as a function of time assuming $\beta=1.9$.
  The shaded region comprises the uncertainty range on $M_{\rm
    ecl,max}$ given by the observational uncertainties on $\sigma_z$
  (upper panel).  The region between the old horizontal dashed lines
  indicates the $M_{\rm ecl,max}$ needed to account for the vertical
  thick-disk kinematics, the thick disk having been born between about
  10 and 12~Gyr ago.  From Kroupa (2002).}
\label{fig:kr_agesigma}
\end{figure}

An immediate application of this theory can be seen in the hitherto
unsolved problem that the observed velocity dispersion of
solar-neighbourhood stars increases more steeply with age than is
predicted from secular heating of the stellar orbits
(Fig.~\ref{fig:kr_agesigma}).
According to the standard picture, stars are formed with a small
velocity dispersion of about 10~km/s. The velocity dispersion
increases with time because the stars scatter on giant molecular
clouds and spiral density waves, and because the disk mass increases
through accretion. Jenkins (1992) calculated this diffusion of stellar
orbits with time (the solid curve in the upper panel of
Fig.~\ref{fig:kr_agesigma}).  The observed vertical velocity
dispersion data (Fuchs et al. 2001) are plotted in the top panel of
Fig.~\ref{fig:kr_agesigma}.  The more recent work of Nordstr\"om et
al. (2004) shows similar results. It can be seen from
Fig.~\ref{fig:kr_agesigma} that the theory cannot account for the
observed increase of $\sigma_z$. A ``heat source'' appears to be
missing.

Adding popping star clusters may resolve this problem of the missing
heat source.  The theoretical velocity dispersion, $\sigma_z$, can be
calculated readily from ${\cal D}(v_z)$ as a function of $M_{\rm
  ecl,max}$ for various $\beta$ and an assumed small cluster--cluster
velocity dispersion of 5~km/s.  As a first ansatz we thus say that the
clusters form in a very thin MW disk.  Fig.~\ref{fig:kr_agesigma}
indicates what sort of $M_{\rm ecl,max}(t)$ with $\beta=1.9$ would be
needed to explain the observed velocity dispersion assuming Jenkins'
(1992) model applies. The result is that $M_{\rm ecl,max}$ takes
reasonable values and that it decreases towards younger ages.

The interesting suggestion coming from this approach is that the MW
disk may have been quietening down over time as its gas reservoir was
depleted. It may have begun forming with a high SFR reaching to
$M_{\rm ecl,max}\approx 10^6\,M_\odot$ thereby producing the thick
disk about 10--12~Gyr ago. Such a high SFR in the early primarily
gaseous MW disk may have been induced by a tidal perturbation from a
nearby neighbour galaxy (Kroupa, Theis \& Boily 2004). Or it may be
the result of an early disk instability (Noguchi 1999).  A full
solution to the ECMF needed to match the vertical thick-disk
kinematics can be found in Kroupa (2002).
The thin disk may then have begun its assembly from star-cluster
populations with a maximum cluster mass decreasing to the present
time. The decreasing $M_{\rm ecl,max}$ would come about from a
decreasing SFR (Weidner et al. 2004) as the conversion of gas to
stars proceeded.

The above constitutes a simple example of how clustered star-formation
may affect the morphology of galaxies.  The results are very
encouraging, in that the constrained parameters $M_{\rm ecl,max},
\beta$ are reasonable. Combined with the observational evidence
discussed in Section~\ref{sec:kr_birth} the notion that star clusters
are fundamental building blocks of galaxies is sharpened.  Clearly, we
need to calculate the velocity distribution function of expanding
stellar populations, ${\cal D}(v_x, v_y, v_z$), within a realistic and
time-evolving Galactic potential in order to construct
three-dimensional velocity ellipsoids at different locations in the MW
disk and at some given epoch for a pre-scribed star-formation history.
This needs treatment of the diffusion of orbits in the time-evolving
potential. Furthermore, ancient thin-disk clusters with $M_{\rm
  ecl,max}\simgreat 10^6\,M_\odot$ may be evident in the MW disk today
as evolved cluster remnants (eq.~\ref{eq:kr_tev}).

\section{Conclusions}

The measured super-virial velocity dispersion of ONC and R136 stars
suggests that both clusters may be out-of-dynamical equilibrium after
blowing out their residual gas explosively. Explosive gas expulsion
from massive young clusters is also supported by measurements of the
outflow velocities of gas, and the Treasure Chest cluster may be
caught in the act of doing so.  Young clusters containing O~stars
therefore may expand rapidly after gas blow-out loosing a large
fraction of their stars.  This expanding population forms a moving
group (MGI) with a velocity dispersion, $\sigma_{\rm MGI}$, that is
related to the pre-gas expulsion velocity dispersion of stars in the
embedded clusters, and thus to the mass, $M_{\rm cl}$, of the
re-virialised cluster that re-contracts as the nucleus of the
expanding MGI.  GAIA will measure this correlation $\sigma_{\rm MGI} =
{\cal K}_{\rm obs}(M_{\rm cl})$, through which we will get an
important handle on the basic physics acting during cluster formation.

Since most stars form in embedded clusters it follows that cluster
birth may leave non-negligible imprints in the kinematical and
structural properties of galaxies. An example has been presented by
calculating the vertical velocity dispersion of a population of stars
born in an ensemble of star clusters. Relating this dispersion to the
missing heat source in the age--velocity-dispersion relation of solar
neighbourhood stars allows us to estimate the evolution of the
embedded-cluster mass function with time. The result of this
exploratory work is that the MW may have been quietening down
somewhat, that early epochs of the MW disk may have been characterised
by more violent star formation within a thin disk producing clusters
reaching to masses similar to those of present-day globular clusters.
A thick disk may thus be a natural outcome of such events. The merging
of cold-dark matter sub-halos may not be needed to explain the
existence of a thick disk, but satellite mergers are of course not
ruled out as having shaped the old MW disk (Abadi et al. 2003).

A more detailed theoretical study of these processes is needed to
quantify the stellar velocity ellipsoid in dependence of various
evolutionary models of the MW disk, and to construct statistical
models of the MW for future comparison with GAIA data. 

Because this notion that popping clusters may shape entire galaxies is
so important, it will also be urgent to observe the velocity
dispersion and the gas content in as many very young clusters as is
possible. While GAIA will have the defining impact on these ideas, it
will not be able to measure the kinematics of very young stars in
dense clusters, so that ground-based proper-motion and line-of-sight
velocity observations are definitely needed to quantify the dynamical
state of as large a number of very young clusters as is possible.

The beauty of this entire approach lies in the realisation that
small-scale baryonic processes may affect galaxy-scale structures
through stellar dynamics. Star-formation scales may thus impact
cosmological issues in a hitherto unappreciated manner.



\end{document}